\documentclass[twocolumn,epsfig,superscriptaddress]{revtex4-1}

\usepackage{amsfonts}
\usepackage{amsmath}
\usepackage{amssymb}
\usepackage{array}
\usepackage{bbold}
\usepackage{bm}
\usepackage{cancel}
\usepackage{changes}
\usepackage{color}
\usepackage{colortbl}
\usepackage{dcolumn}
\usepackage{dsfont}
\usepackage{enumerate}
\usepackage{float}
\usepackage{flushend}
\usepackage{graphicx}
\usepackage{here}
\usepackage[colorlinks=true,citecolor=blue,urlcolor=blue]{hyperref}
\usepackage[utf8]{inputenc}
\usepackage{mathtools}
\usepackage{multirow}
\usepackage{physics}
\usepackage{ragged2e}
\usepackage{romannum}
\usepackage{stmaryrd}
\usepackage{subfigure}
\usepackage{tabularx}
\usepackage{textcomp}
\usepackage{tikz-cd}
\usepackage{ulem}
\usepackage{url}

\newenvironment{helv}{\fontfamily{phv}\selectfont}{\par}
\DeclareTextFontCommand{\texthelv}{\helv}

\makeatletter
\newcommand{\thickhline}{%
    \noalign {\ifnum 0=`}\fi \hrule height 1pt
    \futurelet \reserved@a \@xhline
}
\newcolumntype{"}{@{\hskip\tabcolsep\vrule width 1pt\hskip\tabcolsep}}
\makeatother

\begin{document}

\draft
\title{Exceptional Classifications of Non-Hermitian Systems}
\author{Jung-Wan Ryu}
\thanks{These two authors contributed equally}
\address{Center for Theoretical Physics of Complex Systems, Institute for Basic Science (IBS), Daejeon 34126, Republic of Korea.}

\author{Jae-Ho Han}
\thanks{These two authors contributed equally}
\address{Center for Theoretical Physics of Complex Systems, Institute for Basic Science (IBS), Daejeon 34126, Republic of Korea.}

\author{Chang-Hwan Yi}
\address{Center for Theoretical Physics of Complex Systems, Institute for Basic Science (IBS), Daejeon 34126, Republic of Korea.}

\author{Moon Jip Park}
\email{moonjippark@hanyang.ac.kr}
\address{Department of Physics, Hanyang University, Seoul 04763, Republic of Korea}
\author{Hee Chul Park}
\email{hc2725@gmail.com}
\address{Department of Physics, Pukyong National University, Busan 48513, Republic of Korea}
\date{\today}

\begin{abstract}
Eigenstate coalescence in non-Hermitian systems is widely observed in diverse scientific domains encompassing optics and open quantum systems. Recent investigations have revealed that adiabatic encircling of exceptional points (EPs) leads to a nontrivial Berry phase in addition to an exchange of eigenstates. Based on these phenomena, we propose in this work an exhaustive classification framework for EPs in non-Hermitian physical systems. In contrast to previous classifications that only incorporate the eigenstate exchange effect, our proposed classification gives rise to finer $\mathbb{Z}_2$ classifications depending on the presence of a $\pi$ Berry phase after the encircling of the EPs. Moreover, by mapping arbitrary one-dimensional systems to the adiabatic encircling of EPs, we can classify one-dimensional non-Hermitian systems characterized by topological phase transitions involving EPs. Applying our exceptional classification to various one-dimensional models, such as the non-reciprocal Su--Schrieffer--Heeger (SSH) model, we exhibit the potential for enhancing the understanding of topological phases in non-Hermitian systems. Additionally, we address exceptional bulk-boundary correspondence and the emergence of distinct topological boundary modes in non-Hermitian systems.
\end{abstract}

\maketitle

\date{\today}

\textcolor{cyan}{\it Introduction} -- Non-Hermitian physical systems exhibit a unique type of singularity, known as exceptional points (EPs), where distinct eigenstates of the Hamiltonian coalesce with each other \cite{Kat66, Hei90, Hei04, Dop16, Che20, Tan20, Yan21}. EPs have garnered significant interest due to their potential applications in diverse fields such as optics, acoustics, and open quantum systems \cite{Xu16, Shi16, Mia16, Din16, Che17, Yan18, Ozd19, Mir19}. As a system undergoes an adiabatic deformation that encircles an EP or EPs, the eigenstates exchange in a nontrivial manner. This eigenstate switching effect allows for the classification of EPs based on the conjugacy class of the permutation group \cite{Ryu22}.

Concurrently, the study of Hermitian topological phases has focused on the classification of topological invariants associated with the Berry phase \cite{Tho82, She03, Has10, Qi11}. The quantized Zak phases in the Su--Schrieffer--Heeger (SSH) model are one representative example \cite{Su79, Asb16, Zak89, Ata13, Xia14, Xia15}. In EPs, the wave functions can accompany an additional geometric (Berry) phase shift of $\pi$ after the encircling of an EP. This connection between EPs and nontrivial Berry phases has been both theoretically and experimentally explored \cite{Hei99, Dem01, Dem04, Lee12, Gao15} and suggests a more complex structure for EPs. As we show below, these two seemingly unrelated phenomena---namely the nontrivial Berry phase and EPs---are intimately tied together.

In this work, we demonstrate that the EPs are characterized by \textit{exceptional classifications}, a scheme we propose to incorporate the information of both eigenstate switching and the additional Berry phase. In addition, by viewing one-dimensional (1D) systems as adiabatic deformations encircling EPs, we achieve a full characterization of 1D non-Hermitian topological systems. This characteristic reveals topological phase transitions between different phases, where the phase transitions accompany the EPs. Our identification of this exceptional class lays the groundwork for further exploration of the rich physics of non-Hermitian systems.

\textcolor{cyan}{\it Classification scheme} -- We consider a general $N$-state non-Hermitian system with two external parameters. When encircling the system's EPs through adiabatic deformation, the eigenstates exhibit the exchange effect, which can be represented by a permutation of the $N$ states \cite{Ryu12, Zho18}. The cyclic structure of such permutations can be formally mapped by the conjugacy class $[\sigma]$ (with representative permutation $\sigma$), which forms a product of cycles. Specifically, we can use the following notation to represent the permutation properties of the EPs, as in \cite{Ryu22, Ham62}:
\begin{eqnarray}
[\sigma] = 1^{n_1} 2^{n_2} \cdots N^{n_N}, \ \ \ \sum_{q=1}^N q n_q = N.
\end{eqnarray}
Each cycle is represented in the form $c^{n_c}$, where $c$ indicates the cycle length (number of encirclings required to return to the initial state), and the superscript $n_c \in \{0,1,\cdots,N\}$ denotes the number of $c$-cycles in $[\sigma]$. For instance, in a two-state system, there exist two possible exchanges of eigenstates after the encircling of adiabatic deformations, represented by the conjugacy classes $[e] = 1^2, \ \ \ [\sigma] = 2^1,$ where $e$ represents the identity permutation and $\sigma$ denotes a transposition. 

Furthermore, in addition to the conjugate classification there exist Berry phases of the wave functions. The complex Berry phase \cite{Lia13, Gar88, Dat90, Mos99} can be defined as 
\begin{equation}
    \gamma = i \oint_{\mathcal C} \frac{\left< \phi (\lambda) | \partial_\lambda \psi (\lambda)\right>}{\left< \phi (\lambda) | \psi (\lambda)\right>} d\lambda.
\label{zakphase}
\end{equation}
Here, $\mathcal C$ is a closed path in $\mathcal M \times \mathcal R$ where $\mathcal M$ is the parameter space and $\mathcal R$ is the complex Riemann surface of the energy, $\lambda$ is a parameterization of the path $\mathcal C$, and $\phi$ and $\psi$ respectively are the left and right eigenstates of the Hamiltonian $H(\lambda)$. Due to the complex nature of the energy of non-Hermitian Hamiltonians, we will concentrate on a two-dimensional parameter space (or, codimension two). Since the double encircling of a single EP in parameter space induces a nontrivial $\pi$ Berry phase for the states (see Section I in the Supplemental Material (SM)), the conjugate classifications of the EPs are further sub-classified depending on the presence of the Berry phase. We refer to these finer classifications of the conjugacy class as the \textit{exceptional class}. In the following discussion, we use the notation $\bar c$ for the $c$-cycles with $\pi$ Berry phases.

A constraint arises from the consistency between the switching effect and the Berry phase: the sum of the Berry phases of the cycles in the conjugacy classes should be $0$ and $\pi$ for even and odd permutations, respectively \cite{Lee12}. Note that the parity of permutations remains invariant under conjugation. As an example, consider a two-state system having two conjugacy classes, $[e] = 1^2$, $[\sigma]=2^1$. Under the consistency constraint, the exceptional classes are
\begin{eqnarray}
[e] = 1^2, \bar 1^2, \ \ \
[\sigma] =\bar 2^1,
\end{eqnarray}
where the sums of the Berry phases are $0$ (mod $2\pi$) for $1^2$ and $\bar 1^2$, and $\pi$ for $\bar 2^1$. This is consistent with the parities of $e$ (even) and $\sigma$ (odd). See Fig. \ref{fig01}. Note that $\bar 1^2$ can only appear in systems with multiple EPs. This classification scheme can be generalized to $N$-state systems. Using signed holonomy matrices, the classes can be obtained systematically. We note that some conjugacy classes are connected by gauge transformation and should be identified; details are in Section II of SM.

\textcolor{cyan}{\it Non-reciprocal SSH model} -- In the following, we apply the exceptional classification framework to 1D systems. As an example, we consider the non-reciprocal Su--Schrieffer--Heeger (SSH) model \cite{Lie18, Neh22, Yao18, Gha20, Wan22}, where the Hamiltonian is given as
\begin{eqnarray}
H=\sum_{i} 
te^{\theta} a^\dagger_{i+1}b_{i}+te^{-\theta} b^\dagger_{i-1}a_{i}
+d a^\dagger_{i}b_{i}+ d b^\dagger_{i}a_{i}.
\end{eqnarray}
Here, $d$ represents intra-unit-cell hopping, $t$ denotes inter-unit-cell hopping, and $\theta$ describes the non-reciprocity between left and right directional hopping [Fig. \ref{fig02}(a)]. The energy is measured in $d$, or $d=1$.

We observe three distinct topological phases, where each phase corresponds to adiabatic deformations with different exceptional classifications. Circular strips in Fig. \ref{fig02}(b) represent the corresponding band structure in the complex energy plane. While each complex energy band forms a closed loop in the complex energy space, we observe that the total number of non-separable energy bands varies from a single band (phase II) with a point gap to two bands separated by a line gap (phase I, III) \cite{Kaw19, PhysRevLett.123.066405}. The spectral flow as a function of momentum shows that the bands in phases I and III form a $2\pi$ periodicity (1-cycle band, [$e$]). In contrast, the bands in phase II form a $4\pi$ periodic spectral flow (2-cycle band, [$\sigma$]) due to the eigenstate switching effect. These are represented schematically in Fig. \ref{fig02}(b), where the red and blue strips in phases I and III appear as two separated bands, whereas the strip with both red and blue parts in phase II appears as a connected band. 

In addition, the two 1-cycle bands in phases I and III are further distinguished by the presence of a nontrivial Zak phase \cite{Lie18, Vya21, Neh22}. The bands in phase I (III) have a Zak phase of $\theta = 0~(\pi)$, which indicates that the bands belong to the classification $1$ ($\bar 1$). The nontrivial $\pi$ Zak phase doubles the periodicity of the spectral flow, resulting in a $4\pi$ periodic spectral flow, as schematically represented by the twisted strips in phase III of Fig. \ref{fig02}(b).

The topological phase transitions between each exceptional class can be understood by viewing the Bloch Hamiltonian as an adiabatic encircling of the EPs in two-dimensional parameter space $(t,k)$, as shown in Fig. \ref{fig02}(c). In the Hermitian limit ($\theta=0$), a single Dirac point occurs at $(t,k) = (d,\pi)$. The presence of non-reciprocity splits the Dirac point into a pair of EPs at $(t,k) = (de^{\pm\theta},\pi)$. Within this parameter space, the three topological phases correspond to three distinct loops of the spectral flow that contain zero, one, and two EPs, respectively [Fig. \ref{fig02}(c)]. As a result, the adiabatic encircling at the phase transitions must touch the EPs. The phase transitions between distinct exceptional classes accompany the EPs in the energy spectra [see Fig. \ref{fig02}(b)]. 

For phase I where $t < t_-$, the black loop in Fig.\ref{fig02}(c) contains no EP. Correspondingly, each state returns to its initial state after a single loop in the parameter space, consistent with the trivial Zak phase. Thus, systems in phase I is identified as class $1^2$. For phase II where $t_- < t < t_+$, the presence of an EP within the blue loop, a switching of states under spectral flow with an additional $\pi$ Zak phase. Accordingly, phase II is identified as $\bar 2^1$. Finally, for phase III where $t > t_+$, the red loop encloses a pair of EPs that results in the states returning to their initial state after a single encircling, and each state has a Zak phase of $\pi$. Thus phase III is identified as class $\bar 1^2$. Based on the classification scheme introduced in the previous section, this model exhausts all possible classes of two-state systems. Furthermore, when $t>t_+$, there is no switching of states as the loop encircles the two EPs, but the $\pi$ Zak phase is retained, which distinguishes phase III ($\bar 1^2$ class) from phase I ($1^2$ class). As a result, our model exhausts all possible exceptional classes of 1D two-band systems.

\textcolor{cyan}{\it{Application to generic systems}} -- To illustrate the general applicability of our classification, we now consider three-band systems. According to our classification (Table I in SM), there exist five exceptional classes as follows, 
\begin{eqnarray}
    ~[e]=  1^3, \ 1^1 \bar 1^2,
    ~[\sigma] = \bar 1^1 2^1, \ 1^1 \bar 2^1 ,
    ~[\tau] =  3^1.
\end{eqnarray}
Our proposed 1D three-band model, which realizes all possible exceptional classes, has the following Bloch Hamiltonian [see Fig.~\ref{fig03}(a) for the real-space structure]:
\begin{equation}
    h(k) = 
\begin{pmatrix} 
 0 & d_1 & d_2 + t e^{\theta} e^{-i k} \\
 d_1 & 0 & d_1 \\
 d_2 + t e^{-\theta} e^{i k} & d_1 & 0 \\
\end{pmatrix}\ .
\label{eq_3by3}
\end{equation}
The corresponding phase diagram is depicted as a function of $t$ and $\theta$ in Fig. \ref{fig03}(b). In the Hermitian limit ($\theta=0$), two Dirac points occur, where one Dirac point corresponds to the band touching between the second and third bands and the other to the first and second bands, respectively. As non-reciprocity is turned on ($\theta\neq 0$), each Dirac point is split into a pair of EPs (total four EPs), each of which corresponds to a phase transition point in $(t,k)$ space.

First, we consider the case of $\theta = 0.1$. The energy band structures are shown as schematic figures in the inset of Fig. \ref{fig03}(b). When $t<t_{r-}$, there are three 1-cycle bands with zero Zak phases. When $t_{r-} < t < t_{r+}$, the second and third bands are not separable, while the first band is separable. The non-separable 2-cycle bands have a $\pi$ Zak phase, and the separable 1-cycle band has a zero Zak phase. When $t_{r+} < t < t_{b-}$, the three bands are fully separated. The second and third bands have $\pi$ Zak phases, but the first band has a zero Zak phase. When $t_{b-} < t < t_{b+}$, the sum of the Zak phases of the first and second bands changes from $\pi$ to $0$ at $t=t_{b-}$. The first and second bands are not separable. When $t>t_{b+}$, the three bands are fully separated. The first and third bands have $\pi$ Zak phases, but the second band has a zero Zak phase.

Next, we consider the case of $\theta = 0.4$. When $t'_{b-} < t < t'_{r+}$, the three bands are not separated and the Zak phase of the non-separable 3-cycle band is zero. In Fig.~\ref{fig03}(c), the real band structures are shown for two different phases, $t_{r+}<t<t_{b-}$ for $\theta = 0.1$ (left) and $t'_{b-}<t<t'_{r+}$ for $\theta=0.4$ (right). The topological phases of the band structures are the same as the case of $\theta = 0.1$ in the other regimes. The phase transitions at the EPs are equivalent to adding or removing an EP from the closed loop around the EPs following the adiabatic encircling. The different topological phases result from the positions as well as the number of the EPs in the space ($t$, $k$). In summary, the phases are identified as the classes $1^3$, $1^1 \bar 2^1$, $1^1 \bar 1^2$, $\bar 1^1 2^1$, $1^1 \bar 1^2$, and $3^1$, as shown in Fig.~\ref{fig03}(b). Thus the model exhausts all possible classes of three-state systems.

\textcolor{cyan}{\it Exceptional bulk-boundary correspondence} -- The nontrivial Zak phase manifests as a well-defined topological boundary mode even in non-Hermitian systems. Such boundary modes are distinguished from Hermitian topological boundary modes as the phase transition of the topological phase is manifested by a single EP rather than by a Dirac point. 

In the open boundary condition, the non-Hermitian SSH model exhibits two energy continuum separated by a gap where the gap closes at $t=t_c=1$, signifying a phase transition. See Fig.~\ref{fig04}(a) and (c). For $t>t_c$ (nontrivial phase), mid-gap states emerge [red dots in Fig.~\ref{fig04}(c)] as a consequence of the nontrivial Zak phase. It is noteworthy that in the Hermitian limit, $\theta=0$, these states correspond to edge states. Figure \ref{fig04}(b) and (d) display the phase rigidities of the states{, $r_i = \braket{\phi_i}{\psi_i} / \sqrt{\braket{\phi_i}{\phi_i} \braket{\psi_i}{\psi_i}}$, where $\phi_i$ and $\psi_i$ are left and right eigenstates, respectively, and $i$ is the index of the states \cite{Rot09, Alv18}. The mid-gap states are edge states, and their phase rigidities are significantly smaller. This implies that the edge states are much closer to the EP, which is consistent with the phase transition line at $t=1$.

\textcolor{cyan}{\it Discussion} -- In conclusion, we have introduced a new classification scheme for 1D non-Hermitian topological phases based on exceptional classes. The approach combines the information of both eigenstate switching and the additional Zak phase. We demonstrated the applicability of this classification to both two-band and three-band systems, revealing the rich structure of non-Hermitian topological phases and the interplay between exceptional points and topological properties. We expect that our classification scheme can be applied to other non-Hermitian systems and utilized to reveal the structure of intertwined topological and non-Hermitian properties. Recently, many studies have focused on braid group and knot classifications of non-Hermitian systems \cite{Wan21, Hu21, Li21, Hu22, Woj22, Pat22, Zha23}. These classifications are consistent with ours and provide detailed information on EP structures. However, since it is not easy to discriminate all EPs in a complicated local Riemann surface structure in experiments, a global Riemann surface structure can be universally applied to understand the topological properties of general non-Hermitian systems. For this purpose, permutation operators and groups have sufficient information and thus have been widely used to describe the global Riemann surface topology related to multiple exceptional points \cite{Car09, Ryu12, Dem12, Zho18, Ryu22}. Our classification scheme based on permutation groups completely describes global Riemann surface topology and wavefunction topology.

\section*{acknowledgments}
We acknowledge financial support from the Institute for Basic Science in the Republic of Korea through the project IBS-R024-D1.

\newpage

\begin{figure}
\centering
\includegraphics[width=1.0\columnwidth]{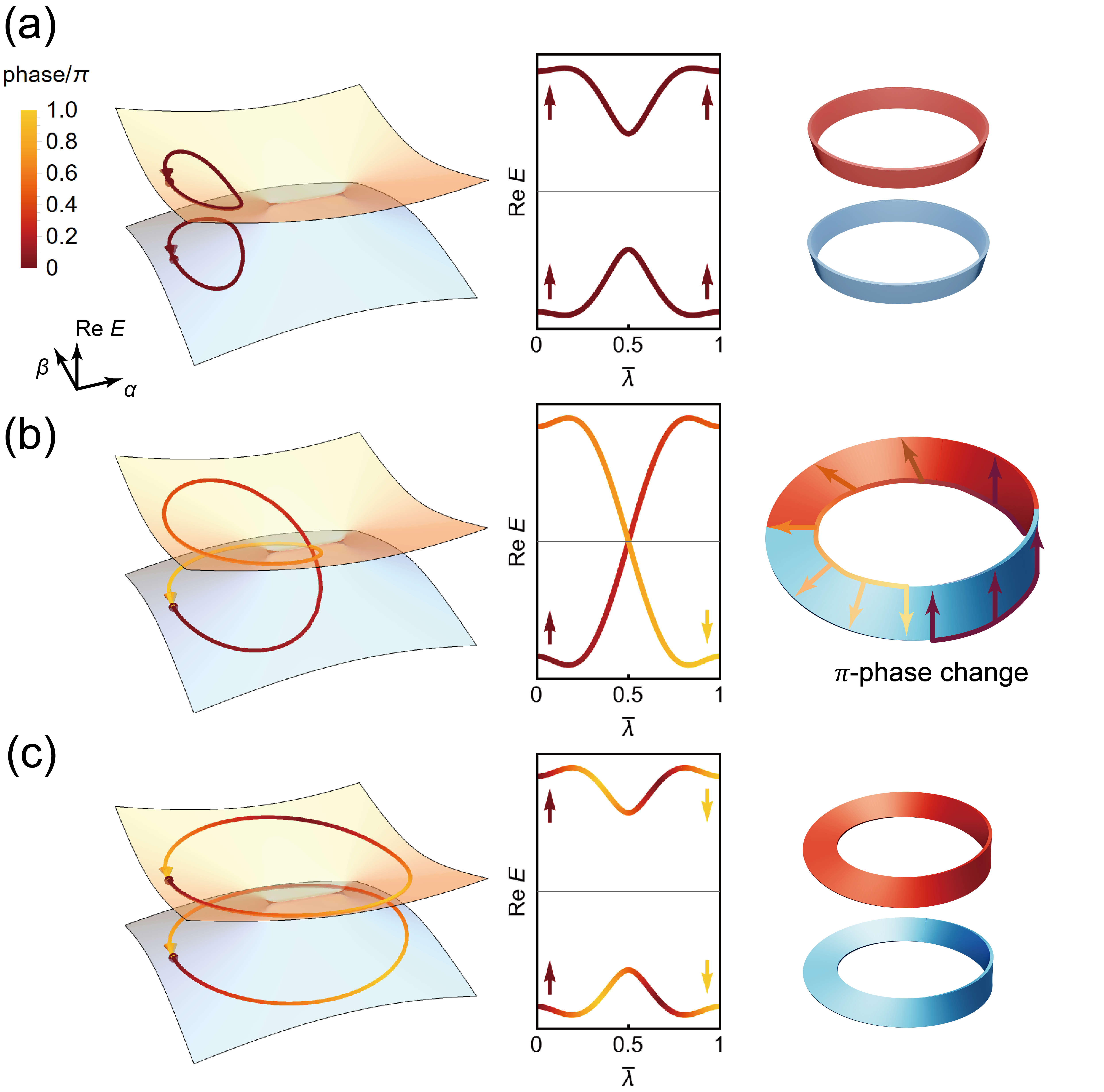}
\caption{Illustration of the switching effect and Berry phase after an encircling of EPs in parameter space, ($\alpha$,$\beta$) space, for three possible cases. The phases of the wavefunctions are shown in the colors of the paths in parameter space (left) and real energy (middle). $\bar\lambda$ is a parameterization of the path projected in ($\alpha$,$\beta$) space. The arrows in the middle column show phase changes after encircling: $0$ for the same directions and $\pi$ for flipped arrows up to multiples of $2\pi$. The schematic figures on the right contain information on both the switching effect and Berry phase (see Section III in SM). The M\"obius strip represents a $\pi$ phase change. (a) Encircling no EPs, no switching, zero phase change. (b) Encircling one EP, state switching after a single encircling, $\pi$ phase change. (c) Encircling two EPs, no switching, $\pi$ phase change.}
\label{fig01}
\end{figure}

\begin{figure}
\centering
\includegraphics[width=1.0\columnwidth]{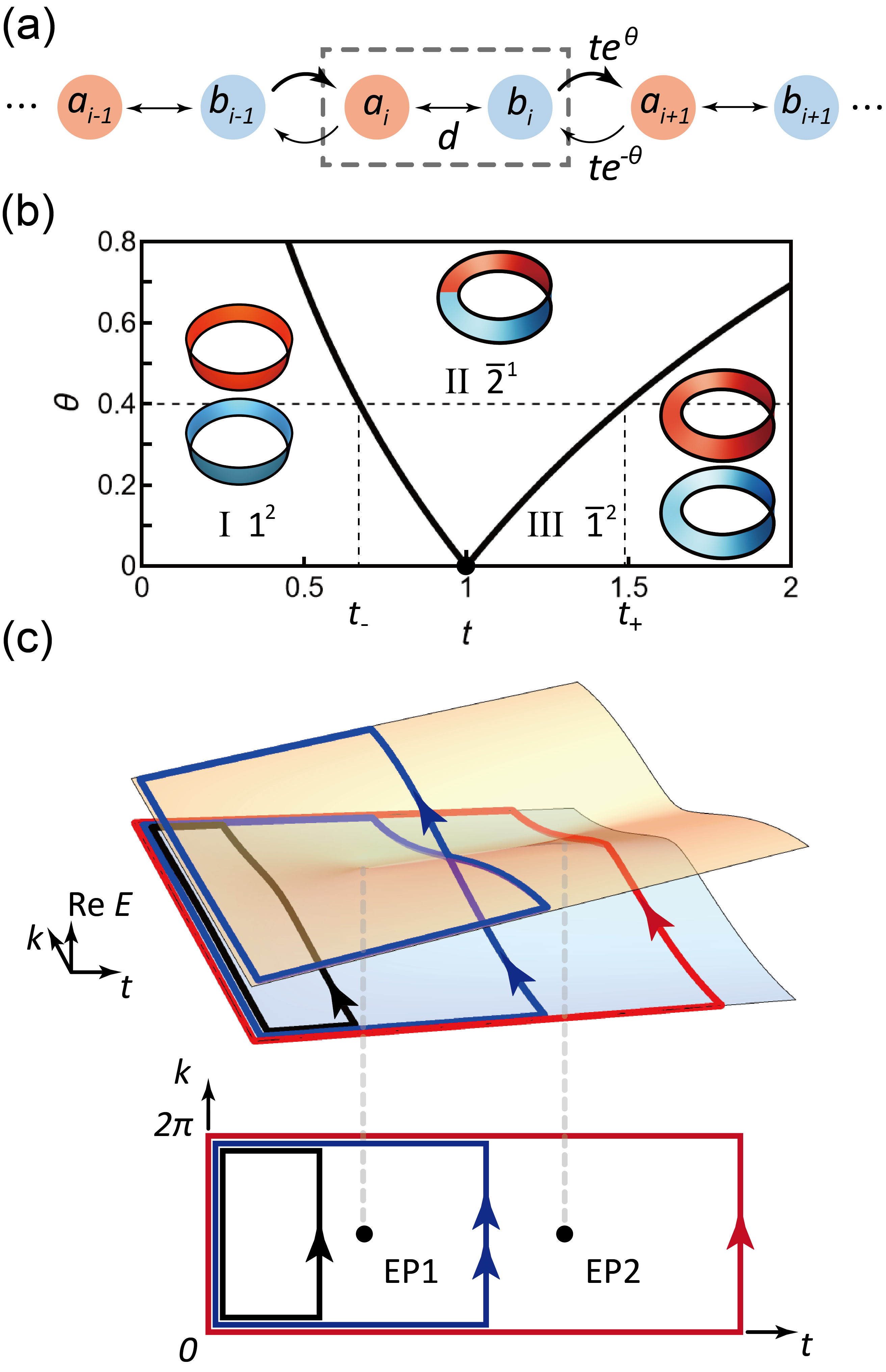}
\caption{(a) Non-Hermitian SSH model with non-reciprocal inter-unit-cell hopping. Intra-unit-cell hoppings are $d$ and inter-unit-cell hoppings are $te^\theta$ and $te^{-\theta}$. The dashed box denotes the unit cell. (b) Phase diagrams of the non-Hermitian SSH model. The black dot denotes a Dirac point, and the curved lines denote EPs. The $1^2$ regime (I) is a trivial line-gapped regime of two trivial $1$-cycle bands, the $\bar{2}^1$ regime (II) is a point-gapped regime of a $2$-cycle non-separable band with a $\pi$ Zak phase, and the $\bar{1}^2$ regime (III) is a line-gapped regime of two $1$-cycle bands with $\pi$ Zak phases. (c) Connection between the Zak phase of the one-dimensional SSH model and the EPs in parameter space, ($t$,$k$) space. Black, blue, and red paths respectively contain zero, one, and two EPs in ($t$,$k$) space, which correspond to $1^2$, $\bar 2^1$, and $\bar 1^2$ phases.}
\label{fig02}
\end{figure}

\begin{figure}
\centering
\includegraphics[width=1.0\columnwidth]{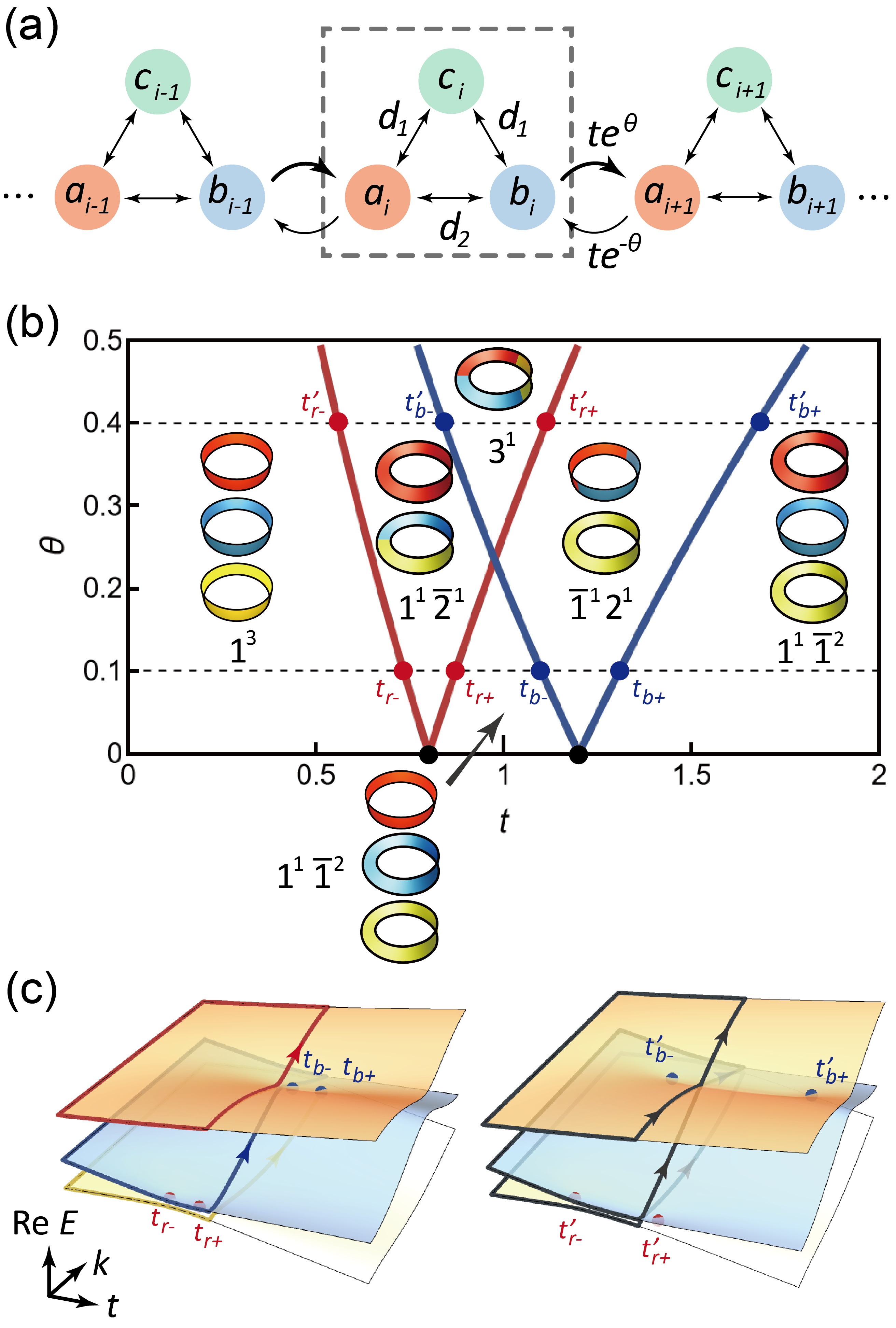}
\caption{(a) Non-Hermitian three-band model with non-reciprocal inter-unit-cell hopping. Intra-unit-cell hoppings are $d_1$ and $d_2$ and inter-unit-cell hoppings are $te^\theta$ and $te^{-\theta}$. The dashed box denotes the unit cell. (b) Phase diagrams of the Hamiltonian, Eq.~(\ref{eq_3by3}), on the ($t$, $\theta$) plane when $d_1=1.0$ and $d_2=0.2$. Black dots represent Dirac points. Red and blue lines represent two pairs of EPs associated with the second and third bands and with the first and second bands, respectively. While there are two kinds of phases at the Hermitian limit ($\theta=0$), $1^3$ and $1^1 \bar{1}^2$, for the non-Hermitian case ($\theta \neq 0$) there are five phases, $1^3$, $1^1 \bar{1}^2$, $1^1 \bar{2}^1$, $\bar{1}^1 2^1$, and $3^{1}$, of which boundaries are pairs of EPs. (c) Real energy surfaces with $\theta=0.1$ (left) and with $\theta = 0.4$ (right) in parameter space, ($t$,$k$) space. The paths are along the Brillouin zone ($k \in [0,2\pi]$) and $t=0,1$. EPs are denoted by red and blue dots, corresponding to the dots shown in (b). There are three separate paths in the $1^1 \bar 1^2$ phase. Blue and yellow paths contain two EPs, resulting in a $\pi$ phase change. In the $3^1$ phase, there is a single path of the 3-cycle band containing two EPs. Three encirclings result in zero phase change.}
\label{fig03}
\end{figure}

\begin{figure}
\includegraphics[width=1.0\columnwidth]{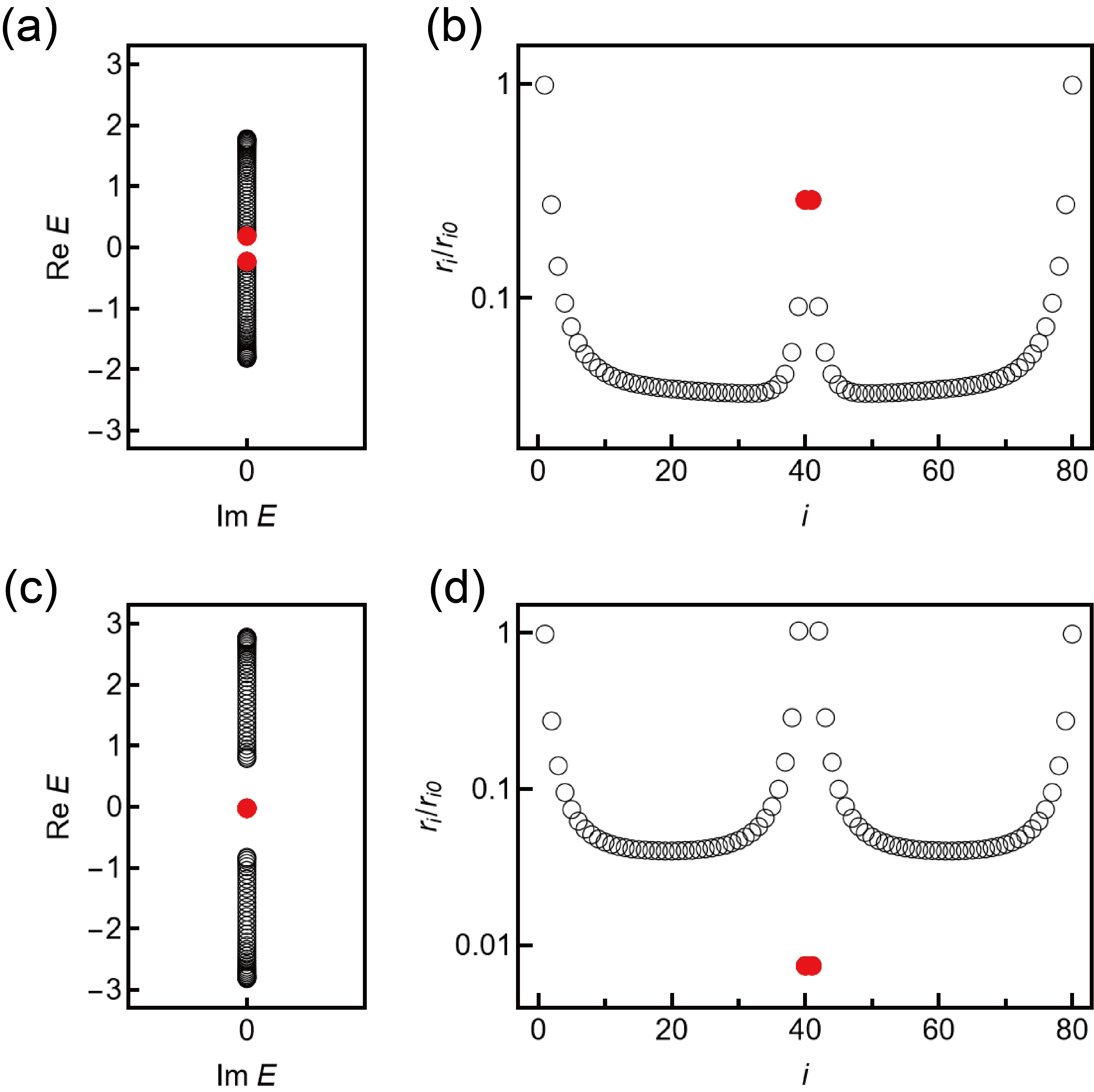}
\caption{(a,c) Complex energy spectra and (b,d) phase rigidities of the eigenstates for the non-Hermitian SSH model under open boundary conditions. (a) and (b) are in the trivial phase, $t=0.8 < t_c=1$, while (c) and (d) are in the nontrivial phase, $t=1.6>t_c$ with $\theta=0.4$. The total number of lattice sites is $2N=80$, and the eigenstates are labeled in increasing order of energy. The red dots are $N$th and $(N+1)$th states, which are the states nearest to the gap in the trivial phase and the edge states in the nontrivial phase.}
\label{fig04}
\end{figure}

\end{document}